\documentstyle[aps,prl,multicol,rotate]{revtex}

\begin{document}
\input{psfig}
\draft

\title{Is Random Close Packing of Spheres Well Defined?}

\author{S. Torquato$^{(1)}$\protect\cite{torquato}, T. M. Truskett$^{(2)}$, and 
P. G. Debenedetti$^{(2)}$}

\address{$^{(1)}$  Department of Chemistry and Princeton Materials Institute, 
Princeton University, Princeton, NJ 08544}
\address{$^{(2)}$  Department of Chemical Engineering,
Princeton University, Princeton, NJ 08544}

\maketitle

\begin{abstract}
Despite its long history, there are many fundamental
issues concerning random packings of spheres that remain elusive,
including a precise definition of {\em random close packing} (RCP).
We argue that the current picture of RCP cannot be made mathematically precise
and support this conclusion via a molecular dynamics
study of hard spheres using the Lubachevsky-Stillinger compression algorithm.
We suggest that this impasse can be broken by introducing
the new concept of a {\em maximally random jammed} state, 
which can be made precise.
\end{abstract}


\begin{multicols}{2}

Random packings of identical
spheres have been studied by biologists, materials
scientists, engineers, chemists and physicists to understand
the structure of living cells, liquids, granular media, glasses
and amorphous solids, to mention but a few examples.
The prevailing notion of random close packing (RCP) is that it is the maximum
density that a large, random collection of spheres can attain
and that this density is a universal quantity.
An anonymous author summarizes this traditional view as
follows: ``ball bearings and similar objects have been shaken,
settled in oil, stuck with paint, kneaded inside rubber
balloons -- and all with no better result than (a packing fraction of) 
 ... $0.636$.''~\cite{An72}

One aim of this paper is to reassess this
commonly held view.  First, we observe that there
exists ample evidence in the literature (in the
form of actual and computer experiments) to 
suggest strongly that the RCP state is ill-defined and, unfortunately, dependent
on the protocol employed to produce the random packing 
as well as other system characteristics.
In a classic experiment, Scott and Kilgour \cite{Sc69} obtained  the 
RCP value $\phi_{c} \approx 0.637$ by pouring
ball bearings into a large container, vertically vibrating
the system for  sufficiently long times to
achieve maximum densification, and extrapolating
the measured volume fractions to eliminate finite-size effects. 
Important dynamical parameters for this experiment
include the pouring rate and both the amplitude and frequency of 
vibration.
The key interactions are interparticle forces, including (ideally)
repulsive hard-sphere interactions, friction between the particles (which
inhibits densification), and gravity. It is clear that the 
final volume fraction can depend sensitively on these
system characteristics. Indeed, in a recent experimental
study \cite{Po97}, it was shown that one can achieve denser
(partially crystalline) packings when the particles are poured at low 
rates into horizontally shaken containers.

Computer algorithms can be used to
generate and study idealized random packings, but the final states are clearly
protocol-dependent. For example, a popular rate-dependent densification 
algorithm~\cite{Jo85} achieves $\phi_c$
between $0.642$ and $0.649$,  a Monte Carlo scheme~\cite{To88}
gives $\phi_c \approx 0.68$, and a ``drop and roll''
algorithm~\cite{Vi72}
yields $\phi_c \approx 0.60$. It is noteworthy
that in contrast to the last algorithm, 
the first two algorithms 
produce configurations in which either the majority or all of the
particles are not in contact with one another.  
We are not aware of any algorithms that truly
account for friction between the spheres.

However, we suggest that the aforementioned inconsistencies and deficiencies
of RCP arise because it is an ill-defined state, explaining why,
to this day, there is no theoretical determination of the RCP density.
This is to be contrasted with the rigor that has been
used very recently to prove that the densest possible
packing fraction $\phi$ for identical spheres is
$\pi /\sqrt{18} \approx 0.7405$, corresponding 
to the close-packed face-centered cubic (FCC) lattice or its stacking
variants~\cite{Ha99}.

The term ``close packed'' implies that the spheres are in contact with one
another with the highest possible coordination number on average.
This is consistent with the view that RCP is the highest possible 
density that a random packing of close packed spheres can possess.
However, the terms ``random'' and ``close packed'' are
at odds with one another. Increasing the degree of coordination,
and thus, the bulk system density, comes at the expense of disorder.
The precise proportion of each of these competing effects
is {\em arbitrary} and therein lies the problem. In what follows,
we supply quantitative evidence of the ill-defined nature of RCP 
via computer simulations,  and we propose a new notion, that of a 
maximally random jammed state.

A precise mathematical definition of the RCP state should
apply to any statistically homogeneous and isotropic
system of identical  spheres (with specified interactions) 
in any space dimension $d$.
Although we discard the term ``close packed'', we must retain
the idea that the particles are in contact with one another, while
maintaining the greatest generality. 
We say that a particle (or a set of contacting particles)
is {\em jammed} if it cannot be translated
while fixing the positions of all of the other particles in the system.
The system itself is jammed if each particle (and each set of 
contacting particles) is jammed \cite{footnote0}.
This definition eliminates 
systems with ``rattlers'' (freely roaming
caged particles) in the infinite-volume limit. We recognize that 
jammed structures created via computer algorithms \cite{BL90}
or actual experiments will contain a very small concentration of 
such rattler particles, the precise concentration of which
is protocol-dependent.  Thus, in practice, one may wish to accommodate this
this type of a jammed structure, although the ideal limit described
above is the precise mathematical definition of a jammed state
that we have in mind.
Nevertheless, it should be emphasized that it is 
the overwhelming majority of spheres that compose the underlying ``jammed'' 
network that confers rigidity to the particle packing.

Our definition of the maximally random jammed (MRJ) state is based on the 
minimization of an order parameter described below.
The most challenging problem is quantifying randomness or
its antithesis: order.
A many-particle system is completely
characterized statistically by the $N$-body probability
density function $P({\bf r}^N)$ associated with finding the
system with configuration ${\bf r}^N$. Such complete
information is never available and, in practice, one
{\it must settle for reduced information}. From this reduced information, 
one can extract a set of scalar {\it order} parameters $\psi_1$, $\psi_2$,
..., $\psi_n$, such that $0 \le \psi_i \le 1$, $ \forall i$, where
$0$ corresponds to the absence of order (maximum disorder)
and $1$ corresponds to maximum order (absence of disorder).
The set of order parameters that one selects
is {\it unavoidably subjective}, given that there is no single
and complete scalar measure of order in the system.

However, within  these {\it necessary limitations}, there is a
systematic way to choose the best order parameters
to be used in the objective function (the quantity
to be minimized). 
The most general objective function  consists of weighted
combinations of order parameters.
The set of all jammed states will define
a certain region in the $n$-dimensional space of order parameters.
In this region of jammed structures, the order parameters can divided up into two
categories:
those that share a common minimum
and those that do not.  The strategy is clear: retain
those order parameters that share a common minimum 
and discard those that do not since they are conflicting measures of order. 
Moreover, since
all of the parameters sharing a common minimum
are essentially equivalent measures of order
(there exists a jammed state in which all
order parameters are minimized), choose
from among these the one that is the most
sensitive measure, which we will simply denote
by $\psi$. From a practical point of view, two order parameters
that are positively correlated will share a common minimum.

Consider all possible configurations of a $d$-dimensional system of identical
spheres, with specified interactions, at a sphere volume fraction
$\phi$ in the infinite-volume limit.
For every $\phi$, there will be a minimum and maximum value
of the order parameter $\psi$.   
By varying $\phi$ between
zero and its maximum value (triangular lattice
for $d=2$ and FCC lattice for $d=3$), the locus of such extrema
define upper and lower bounds within which all structures
of identical spheres must lie.  Figure 1 shows
a schematic (not quantitative) plot of the order parameter versus volume 
fraction.  
Note that at $\phi=0$, the most disordered ($\psi=0$) 
configurations of sphere centers can be realized.  
As the packing fraction is increased, the hard-core interaction prevents
access to the most random configurations of sphere centers (gray region).  
Thus the lower 
boundary of $\psi$, representing the most disordered configurations, increases 
monotonically with $\phi$.
The upper boundary of $\psi$ corresponds
to the most ordered structures at each volume fraction, 
e.g. perfect open lattice structures {\bf ($\psi=1$)}.  
Of course, the details of the  
lower boundary will depend on the particular choice of $\psi$.       
Nevertheless, the salient features of this plot are as  follows: (1)
all sphere structures must lie within the bounds
and (2) the jammed structures
are a special subset of the allowable structures~\cite{footnote1}. 
We define the MRJ state to be the {\it one that minimizes 
$\psi$ among all statistically homogeneous and isotropic jammed
structures}. 

To support the aforementioned
arguments, we have carried out molecular dynamics
simulations using systems of 500 identical hard spheres
with periodic boundary conditions. Starting from an equilibrium
liquid configuration at a volume fraction of $\phi=0.3$,
we compressed the system to a jammed state by the well-known method of 
Lubachevsky and Stillinger \cite{BL90} which allows the diameter
of the particles to grow linearly in time
with a dimensionless rate $\Gamma$. Fig. 2a
shows that the volume fraction of the final jammed states
is inversely proportional to the compression rate $\Gamma$.
A linear extrapolation of the data to the infinite
compression rate limit yields $\phi \approx 0.64$,
which is close to the supposed RCP value reported by Scott
and Kilgour. 

To quantify the order (disorder) in our jammed  structures,
we have chosen to examine two important measures
of order: bond-orientational order
and translational order \cite{footnote2}. The first is obtainable 
in part from the parameter $Q_6$
and the second is obtainable in part
from the radial distribution function $g(r)$
(e.g., from a scattering experiment).
To each nearest-neighbor bond emanating from a sphere, one can associate the
spherical harmonics $Y_{lm}(\theta, \varphi)$, using the bond angles as 
arguments. Then
$Q_6$ is defined by~\cite{PS83}
\begin{equation}
Q_6 \equiv \left(\frac{4 \pi}{13} \sum_{m=-6}^6 
\left|\overline{Y_{6m}}
\right|^2 \right)^{1/2} ,\label{Qldef}
\end{equation}
where $\overline{Y_{6m}}$ denotes an average over all bonds.
For a completely disordered system in the infinite-volume limit,
$Q_6$ equals zero, whereas  $Q_6$ attains its maximum value for 
space-filling structures ($Q_{6}^{FCC}\approx0.575$) in the perfect FCC 
crystal. Thus, $Q_6$ 
provides a measure of FCC crystallite
formation in the system.   For convenience we normalize the orientational
order parameter $Q=Q_{6}/Q_{6}^{FCC}$ by its value in the perfect FCC crystal.

Scalar measures of translational order have not been well studied.
For our purposes, we introduce a translational order parameter
$T$ which measures the degree of spatial ordering, relative to the perfect FCC 
lattice at the same volume fraction. Specifically, 
\begin{equation}
T=\left| \frac{\sum_{i=1}^{N_{C}} (n_{i} - n_{i}^{ideal})}{\sum_{i=1}^{N_{C}}
(n_{i}^{FCC}-n_{i}^{ideal})}
\right|,
\label{T}
\end{equation}
 where $n_{i}$ (for the system of interest) indicates the average occupation
number for the spherical shell of width $a \delta $ 
located at a distance from a reference sphere that equals the $i$th 
nearest-neighbor separation for the open FCC lattice
at that density, $a$ is the first nearest-neighbor distance for that 
FCC lattice, and $N_C$ is the total number of shells (here we choose $\delta=.196$
and $N_{C}=7$).
Similarly, $n_{i}^{ideal}$ and $n_{i}^{FCC}$ are the corresponding shell 
occupation numbers for an ideal gas (spatially uncorrelated
spheres) and the open FCC crystal lattice.  Observe that
$T=0$ for an ideal gas (perfect randomness) and $T=1$ for perfect FCC spatial
ordering.

The relationship between translational and bond-orientational
ordering has heretofore not been characterized. We have measured
both $T$ and $Q$
for the jammed structures generated by the Lubachevsky-Stillinger
algorithm and have plotted the results in the $Q$-$T$ plane
in Fig. 2b \cite{footnote3}. This order plot reveals several key points.
First, we observe that $T$ and $Q$  are positively-correlated and 
therefore are essentially equivalent measures of order for the
jammed structures. 
Therefore, in seeking to determine the MRJ state
using $T$ and $Q$, one would search for jammed structures that 
minimize $Q$, the more 
sensitive of the two measures. Our preliminary results
indicate that the MRJ packing fraction $\phi_{MRJ} \approx 0.64$
for 500 spheres using the Lubachevsky-Stillinger protocol.  It 
should be 
noted, however, that a systematic study of other protocols may indeed 
find jammed states with a lower degree of order
as measured by $Q$. 
Moreover, we notice that the degree
of order increases monotonically with the jammed packing 
fraction~\cite{footnote2}. These results demonstrate that the notion of RCP
as the highest possible density that a random sphere packing can attain
is ill-defined since one can achieve packings with arbitrarily
small increases in volume fraction at the expense of small increases
in order.

For purposes of comparison, we have included in 
the order plot of Fig. 2b results for the equilibrium
hard-sphere system for densities along the
liquid branch and densities along the crystal branch, ending at
the maximum close-packed FCC state \cite{footnote4}.
Interestingly, the equilibrium structures exhibit
the same monotonicity properties as the jammed structures,
i.e., $T$ increases with increasing $Q$ and the degree of order
increases with the packing fraction. Note that neither
$Q$ nor $T$ are equal to unity along the equilibrium
crystal branch because of thermal motion.

To summarize, we have shown that the notion of RCP
is not well-defined mathematically. To replace this 
idea,  we have introduced a new concept:  the maximally random jammed 
state, 
which can be defined precisely once an order parameter $\psi$ is chosen.
This lays the mathematical groundwork for studying randomness in dense 
packings of spheres and initiates the search
for the MRJ state in a quantitative way not possible before.
Nevertheless, significant challenges remain.
First, new and
efficient protocols (both experimental and computational)
that generate jammed states must be developed. Second, since
the characterization of randomness is in its infancy,
the systematic investigation of better order parameters
is crucial.

We thank F. H. Stillinger, T. Spencer, J. H. Conway, and M. Utz
for many valuable discussions. 
S.~T. was supported by the Engineering Research Program of the 
Office of Basic Energy Sciences at the US Department of Energy
(DE-FG02-92ER14275) and the Guggenheim Foundation.  He also thanks the
Institute for Advanced Study in Princeton for the hospitality extended
to him during his stay there. 
P.~G.~D. was supported by the Chemical Sciences
Division of the Office of Basic Energy Sciences at the US Department of Energy
(DE-FG02-87ER13714). T.~M.~T. was supported by NSF.

\end{multicols}
 
\eject
\newpage

\begin{figure}
\centerline {\psfig{file=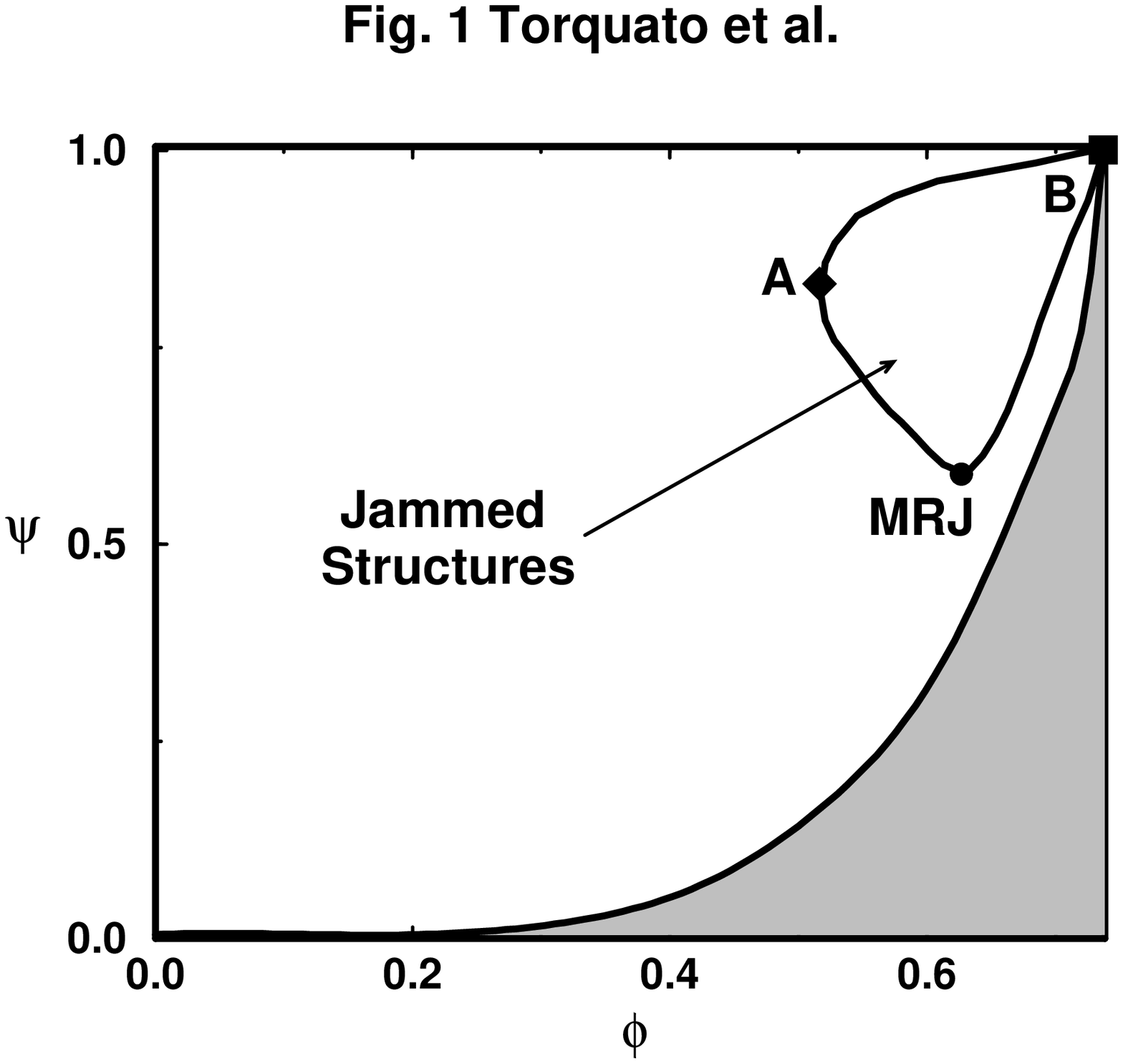}}
\caption{A schematic plot of the order parameter $\psi$ versus volume fraction 
$\phi$ for a system of identical spheres with prescribed
interactions. All structures at a given packing fraction $\phi$, must lie 
between the upper and lower bounds (white region); gray region is inaccessible. 
The boundary containing
the subset of jammed 
structures is shown.   The jammed structures 
are shown to be one connected set; although, in general, they may 
exist as multiply disconnected.  
Point A represents the jammed structure with the 
lowest density and point B  represents the densest ordered jammed structure 
(e.g., close-packed FCC or hexagonal lattice for $d=3$, depending on the 
choice for $\psi$).  The state which minimizes the order
parameter $\psi$ is the maximally random jammed (MRJ) state. }
\end{figure}

\begin{figure}
\centerline {\psfig{file=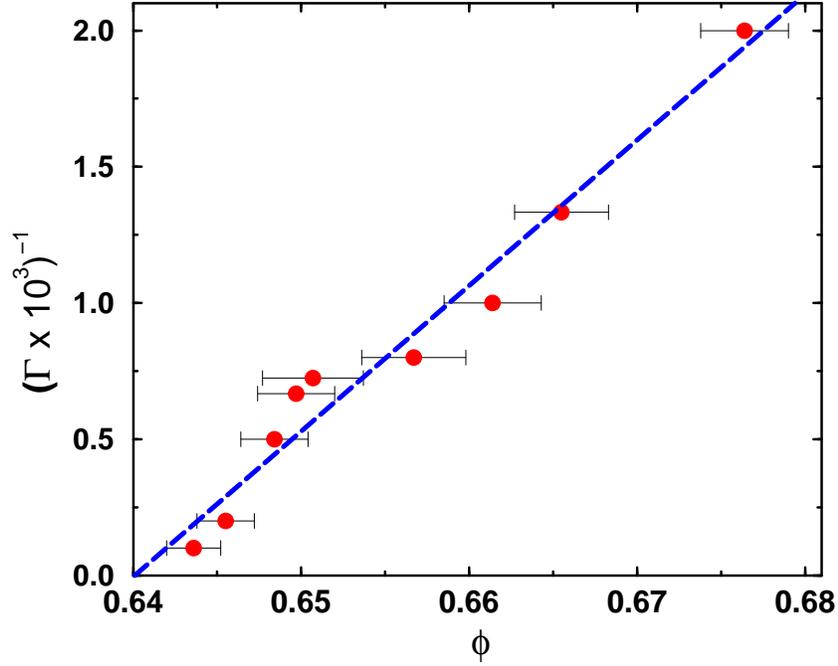,height=4in}} 
\centerline {\psfig{file=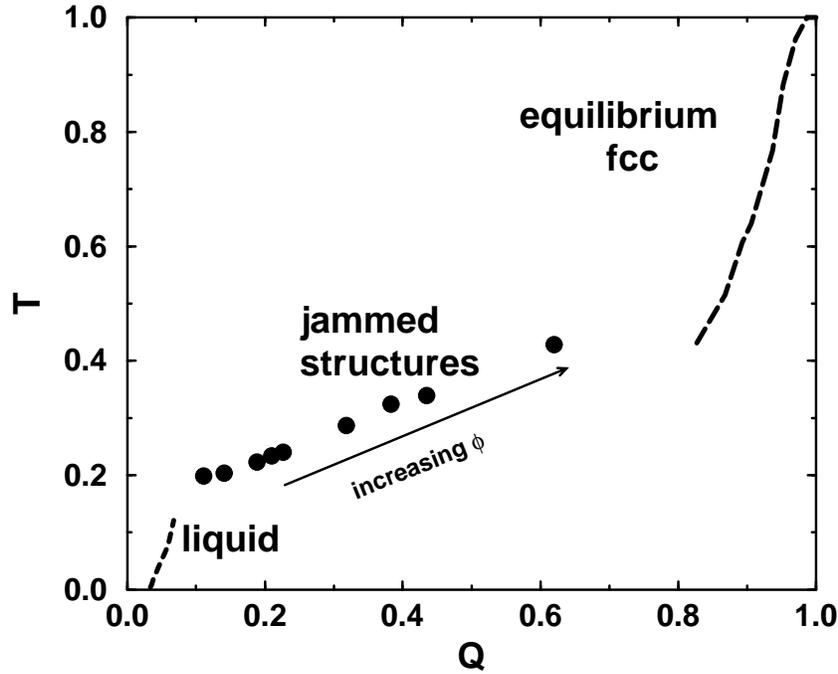,height=4in}} 
\caption{ Molecular dynamics simulation results
for the hard-sphere system.
(a) The reciprocal compression rate $\Gamma^{-1}$ versus the volume 
fraction $\phi$ of the final jammed state of hard spheres
using the molecular dynamics compression protocol of
Lubachevsky and Stillinger \protect\cite{BL90}. 
The jammed state occurs when the 
diameters can no longer increase in time, the sphere collision rate 
diverges, and no further compression can be achieved after relaxing the 
configuration at the jammed volume fraction.  Each point represents the 
average of 27 compressions, and the dashed line is a linear fit to the data,
which yields $\phi \approx 0.64$ when $\Gamma^{-1}=0$.     
(b) The $Q$-$T$ plane
for the hard-sphere system, where $T$ and $Q$ are translational and 
orientational order parameters, respectively.  Shown 
are the average values for the jammed states of Fig. 2a (circles), 
as well as states along the
equilibrium liquid (dotted) and crystal (dashed) branches. }
\end{figure}

\vfill
\eject


\begin{references}
\bibitem{footnote0}
A fascinating open question is: How many jammed configurations exist at
each packing fraction, i.e. what is the density of states for
jammed configurations?
\bibitem{footnote2}
Other reasonable choices
for order parameters were tested, including
an information-theoretic entropy, and resulted in the same qualitative behavior as
the order parameters described
here ($Q$ and $T$).  All of these results, 
as well as the utility of such order parameters for general many-particle systems
(including glasses),
will be reported in a longer paper.
\bibitem{footnote3}
 Our preliminary tests indicate that jammed structures with
packing fractions in the range ($0.68 < \phi < 0.74$) can be
produced when very low compression rates ($\Gamma < 10^{-3}$)
are employed.  However, these structures are
consistently more disordered (as determined by $T$ and $Q$) than
the equilibrium FCC crystal structures at the same packing fraction.
\bibitem{footnote4}
We also found that close-packed crystals composed
of random sequences of FCC and HCP layer placements cannot
be considered to be ``random packings'' by our chosen criteria. 
Both $Q$ and $T$ are not only higher than the minimum
indicated in Fig. 2b for the jammed structures but $Q$ lies in the
range $0.84 < Q <1$, depending on the stacking arrangement.
\end{references}
\end{document}